\documentclass[12pt]{article}

\usepackage{dcolumn}
\usepackage{bm}
\usepackage{amsbsy}		
\usepackage{amsfonts}		
\usepackage{amsmath}		
\usepackage{amssymb}		
\usepackage{slashed}
\usepackage{cite}
\usepackage{url}

\usepackage{psfrag}
\usepackage{subfigure}
\usepackage[latin1]{inputenc}
\usepackage[dvips]{graphicx} 
\usepackage{slashed}

\newcommand{\BR}[2]{\mathrm{BR}\mathinner{(#1\rightarrow #2)}}
\newcommand{\sd}{\mathrm{d}}

\newcommand{\ttbar}{t\bar{t}}

\newcommand{\NAME}[1]{#1}

\newcommand{\BY}[1]{\NAME{#1},}
\newcommand{\IN}[4]{\textit{#1}, \textbf{#2} (#3) #4}

\begin{document}

\begin{center}
\Large\bf\boldmath
\vspace*{0.8cm} New angles on top quark decay to a\\charged Higgs Boson
\unboldmath
\end{center}

\vspace{0.4cm}
\begin{center}
Oscar St\aa l\footnote{Electronic address: {\tt oscar.stal@physics.uu.se}.} \\[0.4cm]
\vspace{0.4cm}
{\sl High-Energy Physics, Dept.~of Physics and Astronomy\\Uppsala University, P.\,O.\,Box 535, SE-751\,21 Uppsala, Sweden}
\end{center}

\vspace{0.3cm}
\begin{abstract}
\noindent 
Top quarks produced in pairs are predicted to experience spin correlations. Due to the large $t\overline{t}$ statistics expected for the LHC, it should be possible to search for new physics effects in angular variables sensitive to these correlations. We investigate, for a general two-Higgs Doublet Model (2HDM), the charged Higgs boson decay of the top quark through the channel $t\to bH^+\to b\tau^+\nu_\tau$. Analytic results are presented on the spin analyzing coefficients for this decay mode. We then explore in some detail the correlation phenomenology in the Type II 2HDM. Finally we present a hadron-level Monte Carlo analysis, illustrating distributions in azimuthal angles which are sensitive to correlations in the transverse plane. These observables are accessible also in the $\tau$ channel, and are therefore particularly interesting for analyzing the $t\to bH^+$ decay.\\
\end{abstract}

\section{Spin correlations in $t\overline{t}$ hadroproduction}
\label{sect:spincorr}
When top quarks are pair-produced in a hadronic collision, the spin projections of the $t$ and the $\overline{t}$ can be correlated to a certain degree when a suitable basis is chosen for spin quantization. There exists an extensive earlier literature on this topic; see for example \cite{ref:SW,ref:MP1,ref:MP2}, or the more recent review \cite{ref:BR2008}. 

The degree of $\ttbar$ spin correlation can be expressed as
\begin{equation}
\label{eq:C}
\mathcal{C}({\hat{\bf a}},{\hat{\bf b}})=4\left<({\bf S}_t \cdot{\hat{\bf a}})({\bf S}_{\overline{t}}\cdot{\hat{\bf b}})\right>,
\end{equation}
where ${\hat{\bf a}},{\hat{\bf b}}$ are interpreted as spin quantization axes for on-shell $t,\overline{t}$. To determine the value of this observable, as a function of the invariant mass of the top pair $m_{t\overline{t}}$, one folds the parton-level correlations $C_{ij}$ of each partonic subprocess with the parton distribution functions (pdfs). Here we work exclusively in the \emph{helicity basis}, defined by ${ \hat{\bf a}}=\hat{\bf p}_t$ and ${\hat{\bf b}}=\hat{\bf p}_{\overline{t}}$, for which the simple threshold expressions $C_{q\overline{q}}=-1/3$ and $C_{gg}=1$ hold. In the UV limit $C_{ij}\rightarrow -1$ due to helicity conservation. From an NLO QCD calculation \cite{ref:BR1}, a total helicity correlation $\mathcal{C}=0.326$ has been determined for the LHC ($\sqrt{s}=14$ TeV). At leading order the corresponding value is $\mathcal{C}=0.319$. These numbers were obtained using $m_t=175$ GeV and CTEQ6.1 pdfs. 

As a special case of the correlation given by eq.~(\ref{eq:C}), it is possible to consider the summed contributions from correlations along three orthogonal axes. This corresponds to
\begin{equation}
\label{eq:D}
\mathcal{D}=4\sum_i\bigl<({\bf S_t} \cdot{\hat{\bf x}_i})({\bf S_{\overline{t}}}\cdot{\hat{\bf x}_i})\bigr>=4\left<{\bf S}_t\cdot {\bf S}_{\overline{t}}\right>.
\end{equation}
At the LHC the expected value for this observable is $\mathcal{D}=-0.219$ $(-0.212)$ at NLO (LO) \cite{ref:BR1}. The ample $\ttbar$ statistics expected with $\sigma^{\mathrm{NLO}}_{\ttbar}\simeq 900$ pb at $\sqrt{s}=14$ TeV will allow the LHC experiments to measure these correlations for the first time. We note in passing that, for these energies, the degree of correlation in the helicity basis can be substantially increased by sacrificing some statistics through a cut on the maximal $m_{t\overline{t}}$ \cite{ref:MP1}. 

\section{Polarized top quark decay and angular correlations}
In the rest frame of a decaying (on-shell) top quark, with the spin directed along the $z$-axis, the differential decay rate can be written \cite{ref:Jezabek}
\begin{equation}
\label{eq:dGamma}
\frac{1}{\Gamma}\frac{\mathrm{d}\Gamma}{\sd\cos\theta_i}=\frac{1}{2}\Bigl(1+\alpha_i\cos\theta_i\Bigr)
\end{equation}
with $\cos\theta_i={ \hat{\bf p}}_i\cdot { \hat{\bf z}}$ for each decay product $i$ with momentum direction $\hat{\bf p}_i$. Each decay product also has a \emph{spin analyzing coefficient} $\alpha_i$, which follows from the Lorentz structure of the $Wtb$ coupling. The $\alpha_i$ are obtained by integrating the polarized decay matrix element over the appropriate phase space. Performing this calculation, one obtains for the SM with $m_t=172.6$ GeV that $\alpha_W=-\alpha_b=0.39$, $\alpha_l=\alpha_{\bar{d}}=1$, and $\alpha_{\nu_l}=\alpha_u=-0.34$. The analytic expressions are summarized in table \ref{tab:alphas}.

 \begin{table}
 \caption{Spin analyzing coefficients $\alpha_i$ for different decay products within the SM and for a charged Higgs decay of the top quark ($t\to bW^+/H^+\to
bl^+\nu_l$), or equivalently ($t\to bW^+/H^+\to b\bar{d}u$).}         
\begin{tabular}{ccc}
   \hline       
         Analyzing & \multicolumn{2}{c}{Coefficient $\alpha_i$ for decay channel} \\
          particle & $W^+~(\omega=m_W^2/m_t^2)$ & $H^+~(\xi=m_{H^+}^2/m_t^2)$ \\
         \hline
          & & \\[-3pt]        
         $b$  & $-\dfrac{1-2\omega}{1+2\omega}$ & $-\dfrac{A^2-B^2}{A^2+B^2}f(\xi,A,B)$ \\[8pt]
         $W^+/H^+$  & $\dfrac{1-2\omega}{1+2\omega}$ & $\dfrac{A^2-B^2}{A^2+B^2}f(\xi,A,B)$ \\[8pt]
         $l^+~(\bar{d})$  & $1$ & $\dfrac{1-\xi^2+2\xi\ln\xi}{(1-\xi)^2}\dfrac{A^2-B^2}{A^2+B^2}f(\xi,A,B)$\\[8pt]
         $\nu_l~(u)$  & $\dfrac{(1-\omega)(1-11\omega-2\omega^2)-12\omega^2\ln\omega}{(1-\omega)^2(1+2\omega)}$ 
         & $-\dfrac{1-\xi^2+2\xi\ln\xi}{(1-\xi)^2}\dfrac{A^2-B^2}{A^2+B^2}f(\xi,A,B)$\\[12pt]
         \hline
\end{tabular}
\label{tab:alphas}
\end{table}

In order to utilize the correlations described in sect.~\ref{sect:spincorr} through decay distributions, one decay product from each top quark are combined into a doubly differential distribution
\begin{equation}
\label{eq:ddiff2}
\frac{1}{N}\frac{\mathrm{d}^2N}{\mathrm{d}\cos\theta_i\,\mathrm{d}\cos\theta_j}
=\frac{1}{4}\Bigl(1+\mathcal{C}\alpha_i\alpha_j\cos\theta_i\cos\theta_j\Bigr).
\end{equation}
Terms linear in the cosines are forbidden by parity invariance. The angles $\theta_{i}(\theta_{j})$ are determined in the respective rest frames of $t(\overline{t})$, as the angles to the directions of motion of $t(\bar{t})$ in the overall $\ttbar$ CM frame. 
In analogy with eq. (\ref{eq:D}), a similar angular distribution in $\cos\theta_{ij}=\hat{\bf p}_i\cdot\hat{\bf p}_j$ is obtained for the $\mathcal{D}$-type correlation
\begin{equation}
\label{eq:diffct}
\frac{1}{N}\frac{\mathrm{d}N}{\mathrm{d}\cos\theta_{ij}}=\frac{1}{2}\Bigl(1+\mathcal{D}\alpha_i\alpha_j\cos\theta_{ij}\Bigr).
\end{equation}
As noted in \cite{ref:Hubaut}, the distribution given by eq.~(\ref{eq:diffct}) is less sensitive to acceptance loss by phase-space cuts than that given by eq.~(\ref{eq:ddiff2}).

It is possible to measure this type of distributions experimentally to obtain information on $\mathcal{C},\mathcal{D}$ (assuming SM decay) or $\alpha_i,\alpha_j$ (when the correlation can be determined in an independent manner). Correlation measurements at the LHC were studied previously for ATLAS \cite{ref:Hubaut}, with an expected $\sim$few percent precision on determining $\mathcal{C}$ and $\mathcal{D}$ in the dilepton channel ($\alpha_l\alpha_l=1$) using $10$ fb$^{-1}$ of data. Such an analysis would not be statistics limited, inviting us to study rare processes related to new physics together with top quark spin correlations. This will be the topic of the next section.

\section{Top quark decay to charged Higgs bosons}
Should the top quark have non-standard couplings, changes can occur either to the production-level correlation, and/or to the decay distributions. A general account of anomalous $Wtb$ couplings in single top production is given in \cite{ref:TOP08:Aguliar-Saavedra}. If they exist, the same couplings are of course relevant to top decay. 

We shall take a different route and discuss instead a charged Higgs decay of the top quark. This mode could become important in two-Higgs Doublet Models (2HDM) when $m_{H^+}<m_t-m_b$. From LEP, the direct limit on the mass of a charged Higgs boson is $m_{H^+}>79.3$ GeV at $95\%$ CL \cite{ref:LEPlimit}, assuming only the decays $H^+\to \tau^+\nu_\tau$ and $H^+\to c\bar{s}$ are possible.
Parametrizing the $H^+$ interactions with fermions as
\begin{equation}
\begin{aligned}
\label{Eq:HiggsL}
\mathcal{L}_{H^\pm f\bar{f}}=&\frac{g_W}{2\sqrt{2}m_W}
\sum_{\substack{\{u,c,t\}\\\{d,s,b\}}}V_{ud}
H^+\bar{u}\Bigl[A\left(1-\gamma_5\right)+B\left(1+\gamma_5\right)\Bigr]d+\\
&+\frac{g_W}{2\sqrt{2}m_W} \sum_{\{e,\mu,\tau\}}H^+C\bar{\nu}
_l\left(1+\gamma_5\right)l+\mathrm{h.c.},
\end{aligned}
\end{equation}
we determine the $\alpha_i$ for the decay $t\to bH^+ \to bl^+\nu_l$ as a function of the couplings $A,B,$ and $C$. The calculation is performed in the narrow-width approximation for on-shell $t(\bar{t})$, assuming independent decay channels. Further details are described in \cite{ref:ttspin}. The results are given in table \ref{tab:alphas}. For the decay through $H^+$, the coefficients are seen to depend on two universal factors. Since $f(\xi,A,B)$ is very close to unity except in the limit $m_{H^+}\to m_t$, when anyway $\BR{t}{bH^+}\to 0$, it will simply be set to $1$. The coupling factor $(A^2-B^2)/(A^2+B^2)$ is important, and will be further discussed below. Our results agree with those in \cite{ref:Korner}, where also $\mathcal{O}(\alpha_s)$ corrections to $\alpha_H$ are calculated.

\begin{figure}
\begin{centering}
\includegraphics[width=0.5\columnwidth, keepaspectratio]{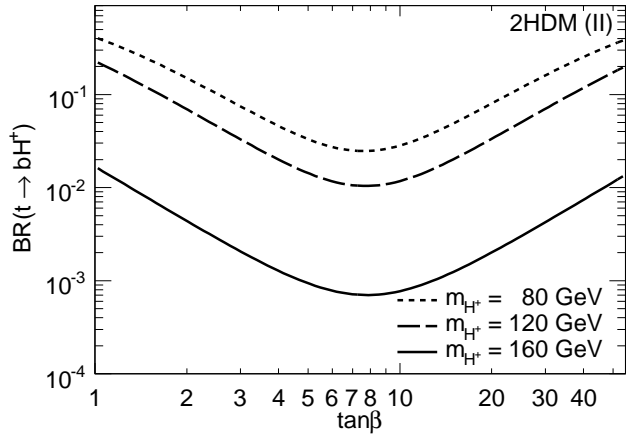}
\caption{Branching fraction for top quark decay $t\to bH^+$ in the 2HDM (II) for different values of $m_{H^+}$.}
\label{fig:BR}
\end{centering}
\end{figure}

Motivated by minimal supersymmetry, the most widely studied 2HDM is the Type II, where one doublet is coupled exclusively to up-type fermions, and the other to down-type fermions. In this model the couplings have values $A=m_u\cot\beta$, $B=m_d\tan\beta$, and $C=m_l\tan\beta$, where $\tan\beta$ is the ratio of the two doublets' vacuum expectation values. Since the Higgs couples proportionally to the fermion masses, we are concerned only with couplings to third generation particles in the following. QCD corrections to the Yukawa couplings are taken into account by renormalizing the fermion masses in the $\overline{\mathrm{MS}}$ scheme at the scale $m_{H^+}$ \cite{ref:Braaten}. Fig.~\ref{fig:BR} shows the resulting branching ratio $t\to bH^+$ as a function of $\tan\beta$ for different $m_{H^+}$.

\begin{figure}
\begin{centering}
\includegraphics[width=0.6\columnwidth, keepaspectratio]{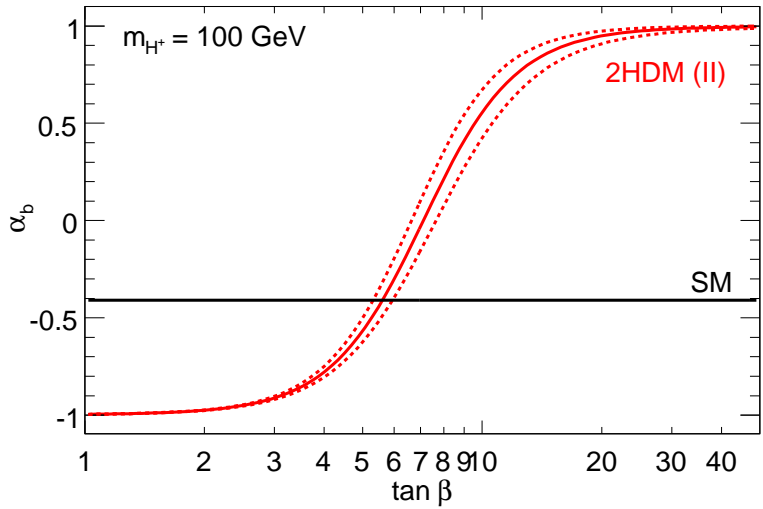}
\caption{Spin analyzing coefficient $\alpha_b$ for top decay in the 2HDM (II) and the SM. Fermion masses in the Higgs couplings are evaluated at the scale $m_{H^+}=100$ GeV. The solid curve shows the pure 2HDM (II) result, while dashed curves correspond to scenarios with large SUSY corrections $\epsilon_b=-\epsilon^{\prime}_t=\pm 0.01$.}
\label{fig:A2B2}
\end{centering}
\end{figure}

In fig.~\ref{fig:A2B2} we illustrate the $\tan\beta$ dependence of $\alpha_b$ for the 2HDM (II).
As seen in the figure, the efficiency to analyze the top quark spin is highest in the limits of small and large $\tan\beta$, where it can reach unity. The $b$ quark (or the $H^+$ itself, since $\alpha_{H^+}=-\alpha_b$) is therefore the most powerful analyzer in the 2HDM. This is a difference compared to the SM, where it is the charged lepton (or, equivalently, the down-type quark from $W\to u\bar{d}$) which carries unit efficiency. 

We consider also SUSY QCD corrections to the $H^+$ couplings, which occur in the 2HDM (II) of the MSSM. The $\tan\beta$-enhanced corrections to the down-type Higgs doublet coupling are studied in \cite{ref:Nierste,ref:Degrassi}. They can be correctly resummed and evaluated to all orders by replacing
\begin{equation*}
\begin{aligned}
 m_b\tan\beta&\to m_b\frac{\tan\beta}{1+\epsilon_b\tan\beta}\\
 m_t\cot\beta&\to m_t\cot\beta(1-\epsilon^{\prime}_t\tan\beta)
\end{aligned}
\end{equation*}
in the couplings. Here $\epsilon_b, \epsilon^{\prime}_t$ are functions which fulfill $|\epsilon_i\tan\beta|<1$. As can be seen in fig.~\ref{fig:A2B2}, the effects on the $\alpha_i$ are small; in the transition region with intermediate $\tan\beta$, they amount to at most a $10-20\%$ correction. The effect cancels completely in the high $\tan\beta$ limit. We note that similar SUSY corrections apply also in the limit of large $\cot\beta$. The conclusions on $\alpha_i$ should be the same for this case. 

\begin{figure}
\begin{centering}
\subfigure{
   \includegraphics[width=0.47\columnwidth,keepaspectratio]{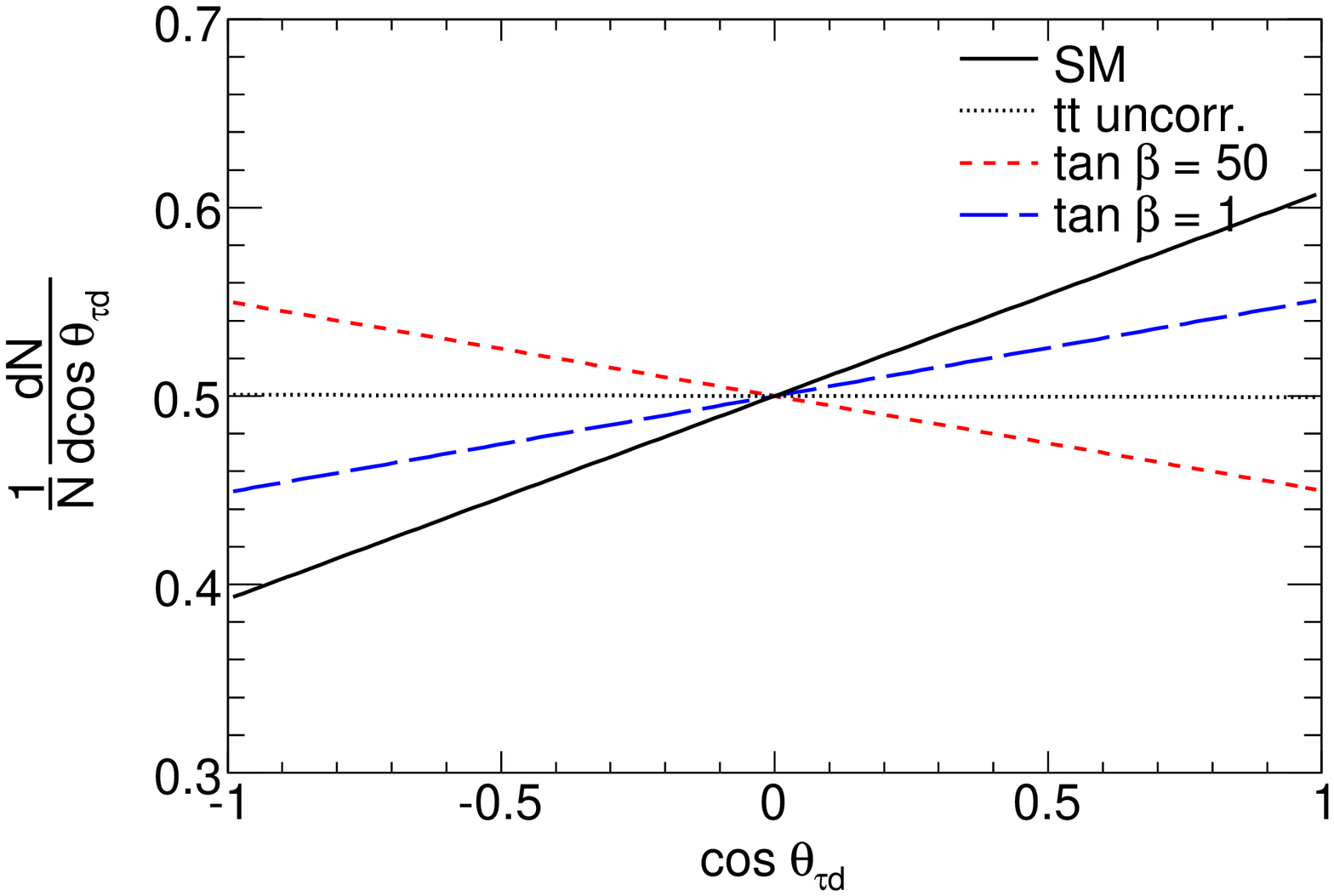}

}
\subfigure{
   \includegraphics[width=0.47\columnwidth,keepaspectratio]{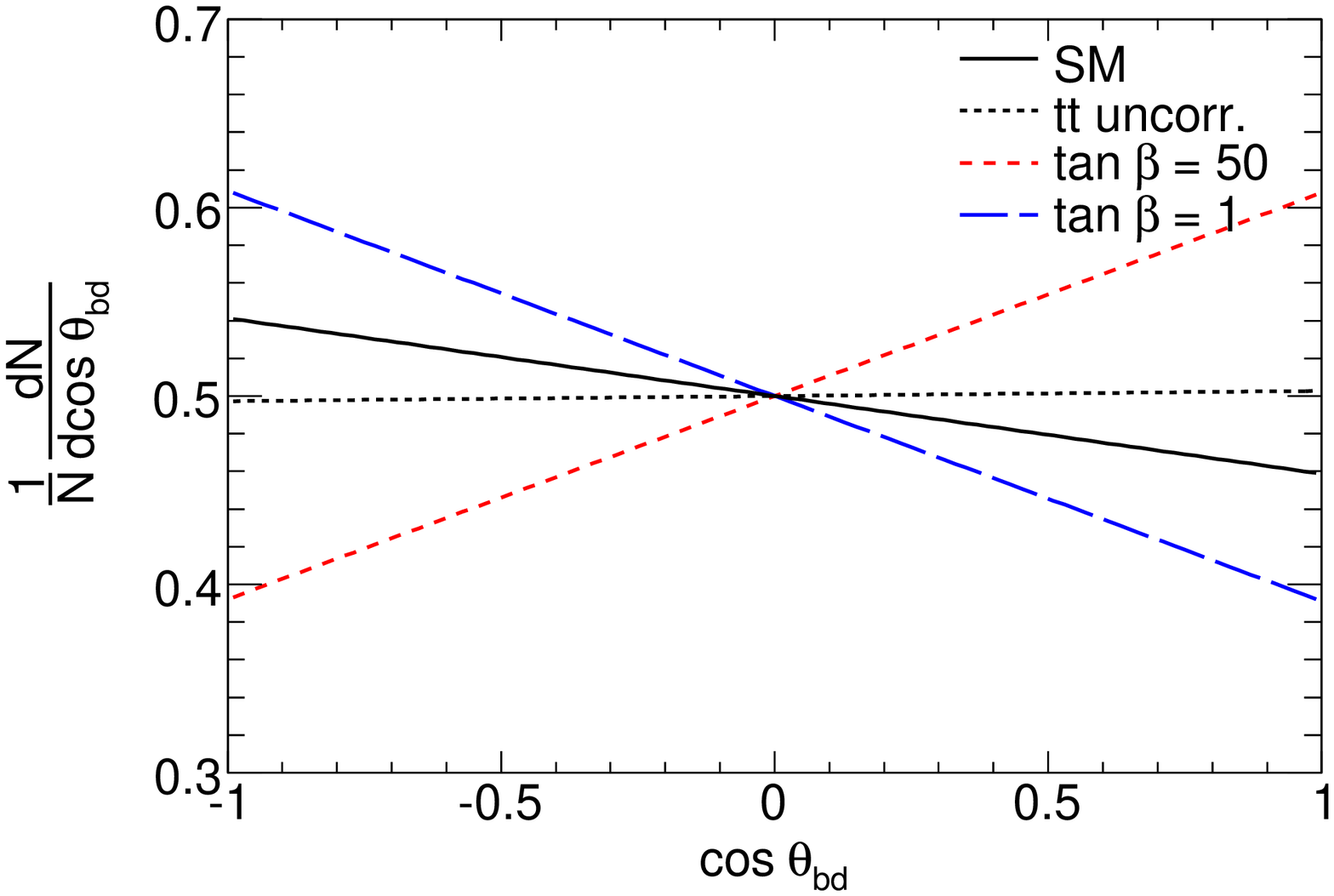}
}
\caption{Truth distributions for different final state particles $(i,j)$. The lines correspond to the SM with (solid black) and without (black dotted) $\ttbar$ spin correlations, the 2HDM (II) for $\tan\beta=1$ (dashed blue), and for $\tan\beta=50$ (dotted red). The results were evaluated using $m_{H^+}=80$~GeV.}
\label{fig:costll}
\end{centering}
\end{figure}
To compare the angular correlations in the SM with those in the 2HDM (II) we display in fig.~\ref{fig:costll} analytic results on the $\mathcal{D}$-type distributions from eq.~(\ref{eq:diffct}). The results are given for the process $pp\to \ttbar\to b\bar{b}\tau^+\nu_\tau\bar{u}d$, where $\tau\nu_\tau$ in the case of 2HDM originates from the decay $H^+\to \tau^+\nu_\tau$ which dominates over a broad $\tan\beta$ interval. The two combinations $(\tau,d)$ (optimal for SM) and $(b,d)$ (optimal for 2HDM) are chosen to illustrate the maximal differences. In the figure the two extreme cases for $\tan\beta$ are shown with the SM expectation, and with a flat distribution corresponding to no $\ttbar$ spin correlations.

\section{Monte Carlo Simulations}
\begin{figure}
\begin{centering}
\subfigure{
   \includegraphics[width=0.47\columnwidth,keepaspectratio]{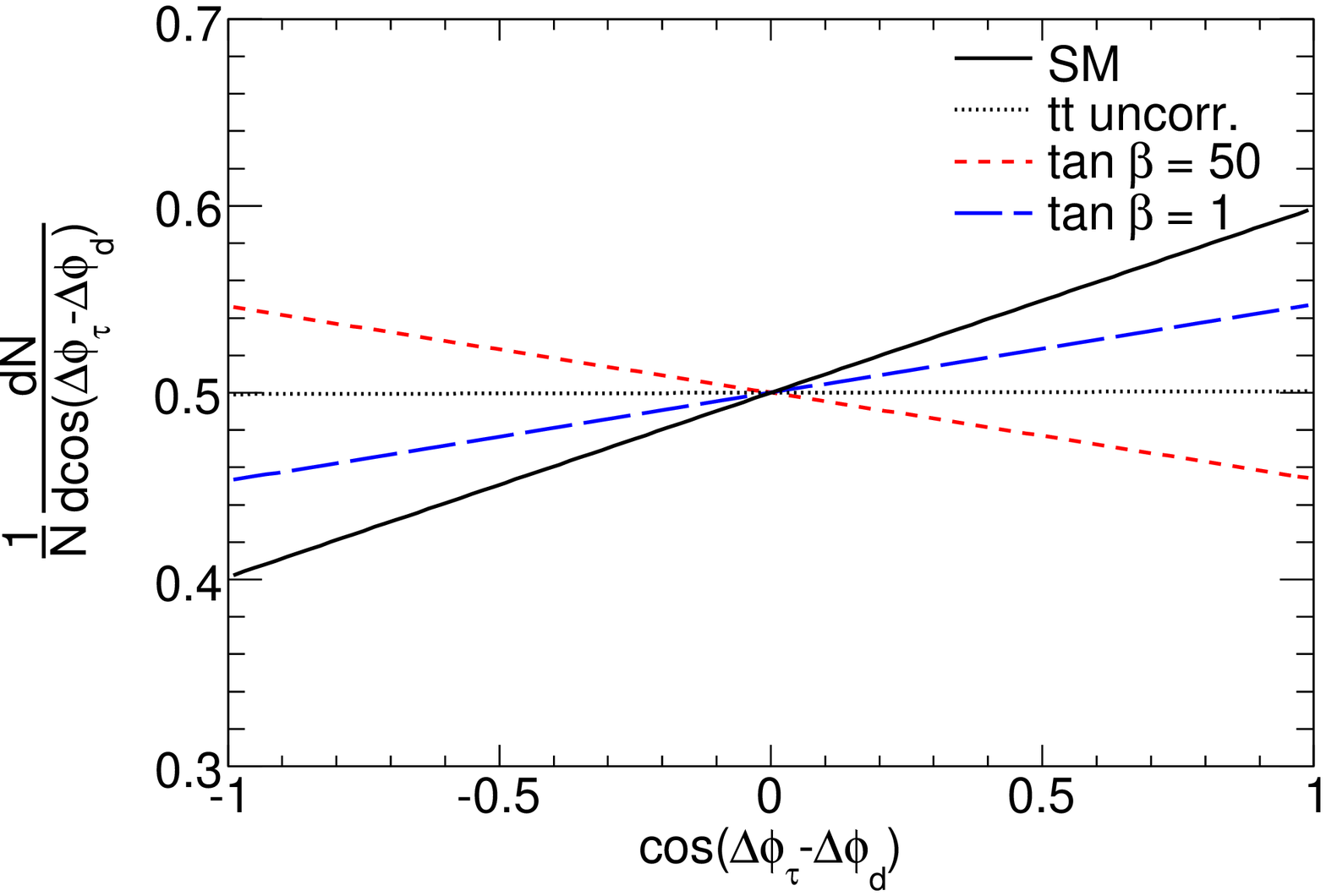}

}
\subfigure{
   \includegraphics[width=0.47\columnwidth,keepaspectratio]{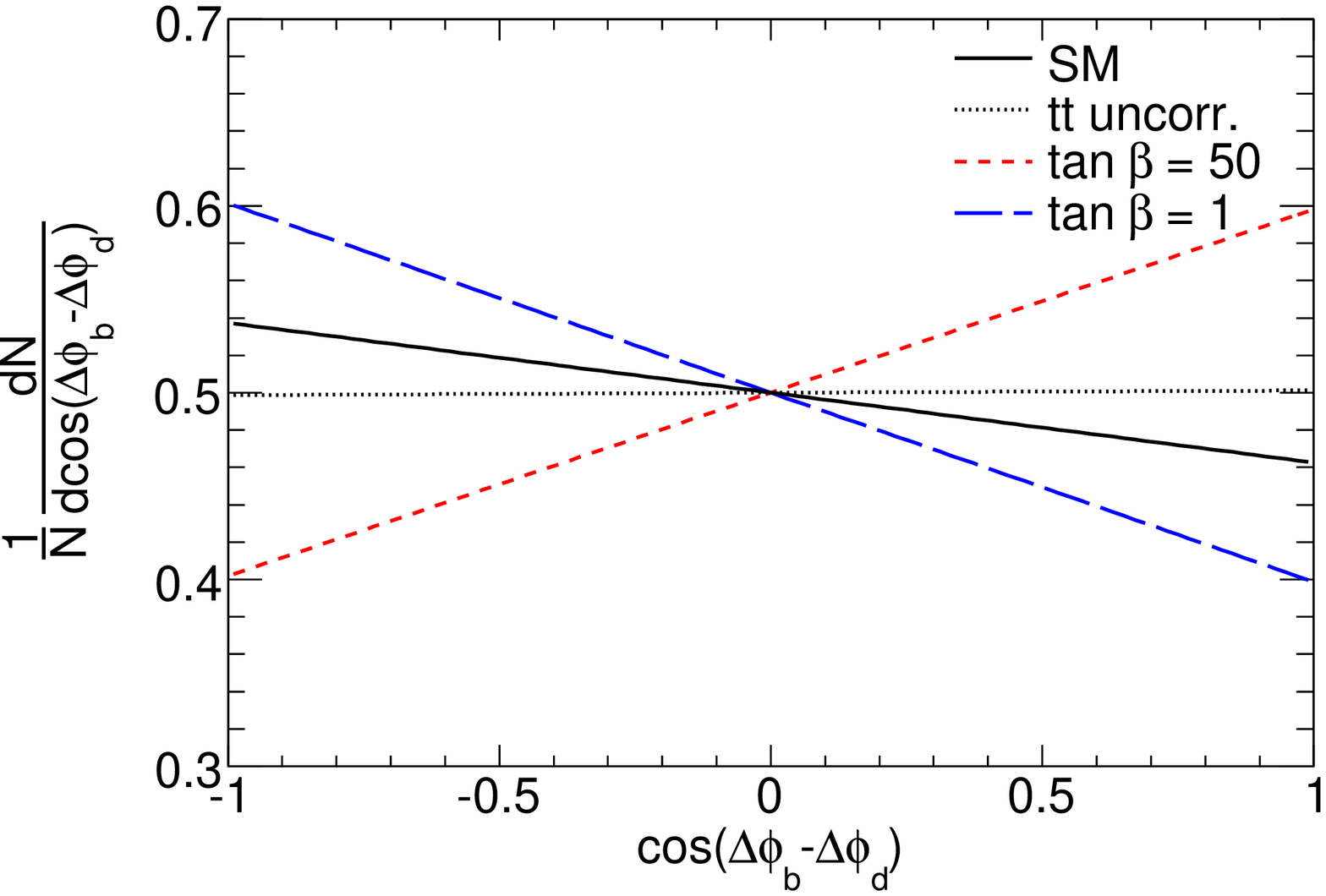}
}
\caption{Azimuthal distributions for different final state particles $(i,j)$. Color coding as in fig.~3. All results for $m_{H^+}=80$ GeV.}
\label{fig:dphill}
\end{centering}
\end{figure}

There are certain limitations in the treatment presented in the previous section, namely that the $\tau$ decay of the $H^\pm$ prevents reconstruction of the $\ttbar$ rest frame since there are too many neutrinos in the final state. This reconstruction is necessary in order to construct the angular distributions discussed above. To address this issue, we consider correlations in the plane transverse to the beam \cite{ref:ttspin}. The angular variable analogous to $\cos\theta_{ij}$ in this case is $\hat{\bf p}_{\perp i}\cdot \hat{\bf p}_{\perp j}=\cos\Delta\phi_{ij}$, where $\hat{\bf p}_{\perp i},\hat{\bf p}_{\perp i}$ are the transverse direction vectors in the transverse rest frames of $t(\bar{t})$. Introducing the notation $\Delta\phi_i$ for the azimuthal angle measured to a fixed axis, for particle $i$ in the transverse rest frame, $\Delta\phi_{ij}=\Delta\phi_{i}-\Delta\phi_{j}$. Using these variables, we expect a distribution of the form
\begin{equation}
\label{Eq:diffdphi}
\frac{1}{N}\frac{\mathrm{d}N}{\mathrm{d}\cos(\Delta\phi_i-\Delta\phi_j)}=\frac{1}{2}\Bigl[1+\mathcal{D'}\alpha_i\alpha_j\cos(\Delta\phi_i-\Delta\phi_j)\Bigr].
\end{equation}
The transverse correlation at the LHC we determine numerically to have the value $\mathcal{D'}=-0.193$ at LO. Comparing figs.~\ref{fig:costll} and \ref{fig:dphill}, their similarity shows that most of the information on the $\ttbar$ spin correlations is indeed accessible in the transverse plane. 

To test if these parton level results apply also at the hadron level we perform a Monte Carlo study of the signal $pp\to \ttbar\to b\bar{b}(H^\pm\to\tau^\pm\nu_\tau)(W^\mp\to q\bar{q})$, with $\tau\to\mathrm{hadrons}$. In addition we generate the same final state for the SM process. The matrix element for the $2\to 6$ process is generated with MadEvent \cite{ref:ME} to preserve the spin correlations. PYTHIA \cite{ref:PYTHIA} is then applied for parton showering, hadronization and generation of the underlying event. The hadronic $\tau$ decay is treated by TAUOLA \cite{ref:TAUOLA}.

A simple spin correlation analysis is employed as follows: We require all particles to have $|\eta|<5$ to account for the detector region. An exclusive mode $k_\perp$ jet finding algorithm is used, with the separator $d_\mathrm{cut}=20$ GeV. Jets with $|\eta|<2.5$, which are matched to a true $b$ quark or $\tau$ lepton, are identified and flavour ``tagged''. The identification of exactly one such $\tau$ jet and two $b$ jets is required. $W$ and $t$ candidates are constructed from the $jj$ and $jjb$ combinations which minimizes the $\chi^2$ difference to the known $W$ and $t$ masses. It is required that the $W$ ($t$) candidates are within $10$ $(15)$ GeV of the true mass values for the events to be retained for further analysis.

Since one of the final state $b$ jets is assigned to the $t$ decaying in the fully hadronic mode (SM), the other $b$ jet can be unambiguously assigned to the prospective $H^+$ side of the event (where also the $\tau$ originates from). We reconstruct the transverse momentum of the $t$ on this side using the combination $p_{\perp}^t=p_{\perp}^{b\mathrm{ jet}}+p_{\perp}^{\tau\mathrm{ jet}}+p_{\perp}^{\mathrm{miss}}$. 

\begin{figure}
\begin{centering}
\subfigure{
   \includegraphics[width=0.47\columnwidth,keepaspectratio]{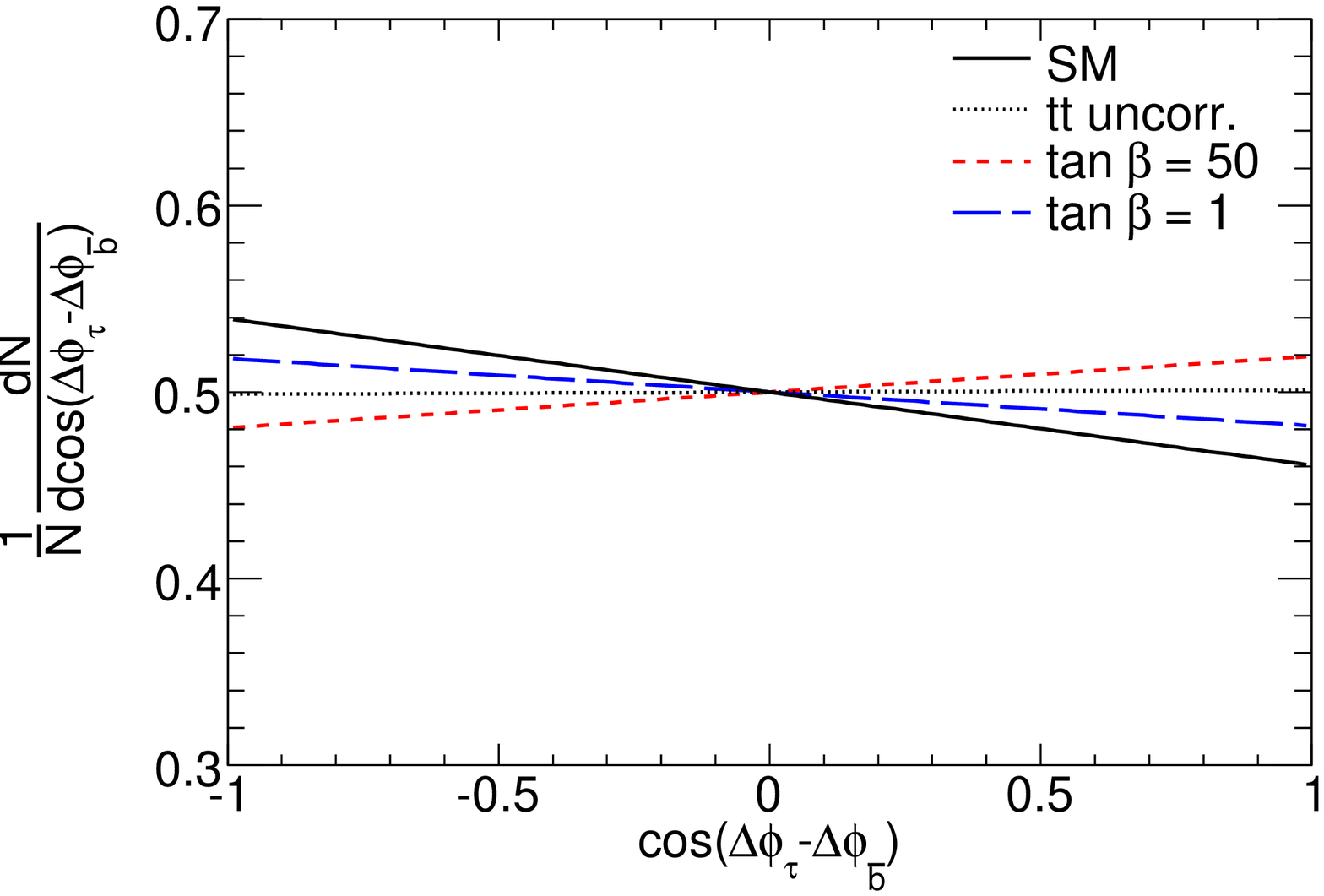}

}
\subfigure{
   \includegraphics[width=0.47\columnwidth,keepaspectratio]{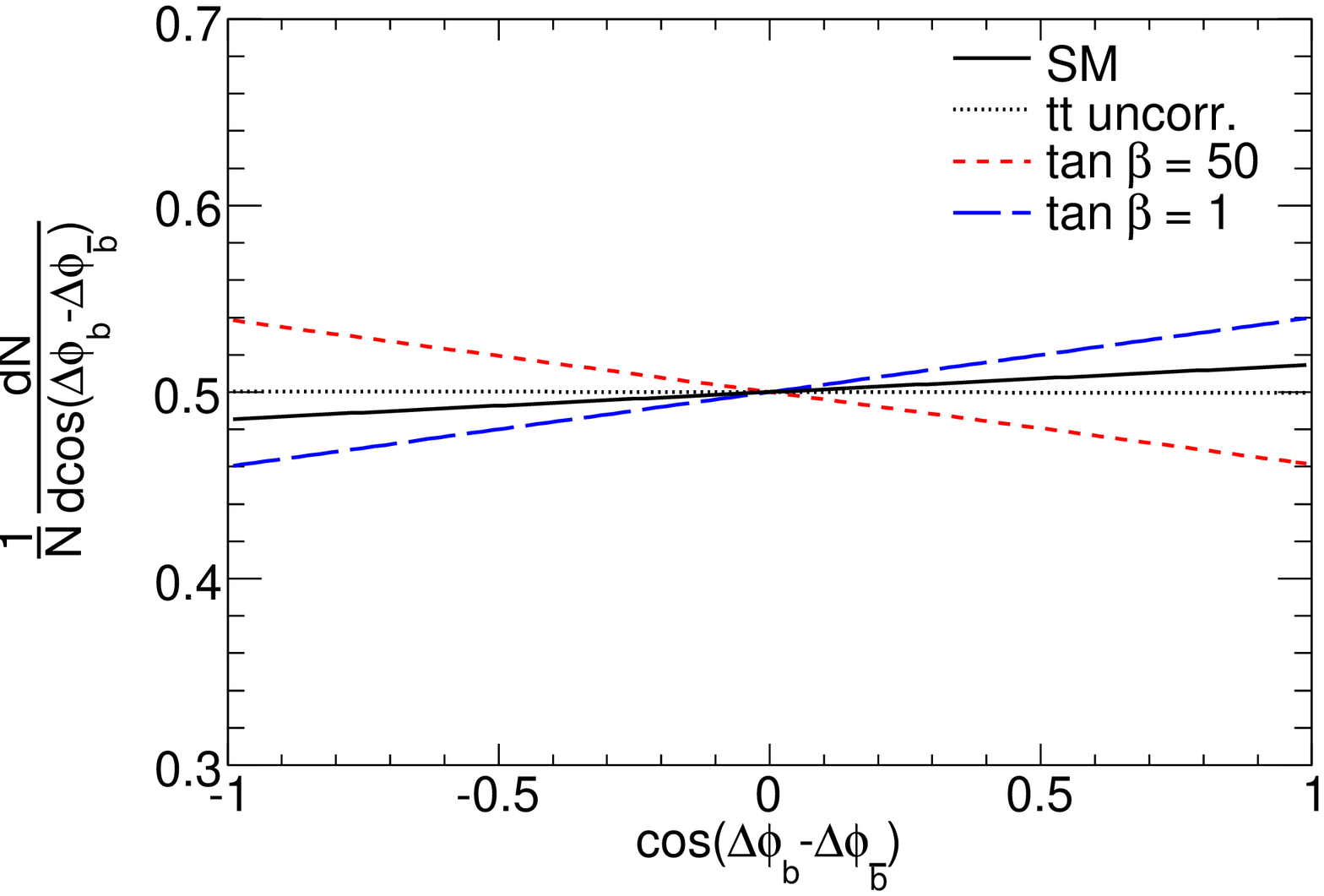}
}
\caption{Truth distributions for different final state particles $(i,j)$. Color coding as in fig.~3. All results were obtained for $m_{H^+}=80$ GeV.}
\label{fig:dphi_leff}
\end{centering}
\end{figure}
\begin{figure}
\begin{centering}
\subfigure{
   \includegraphics[width=0.47\columnwidth,keepaspectratio]{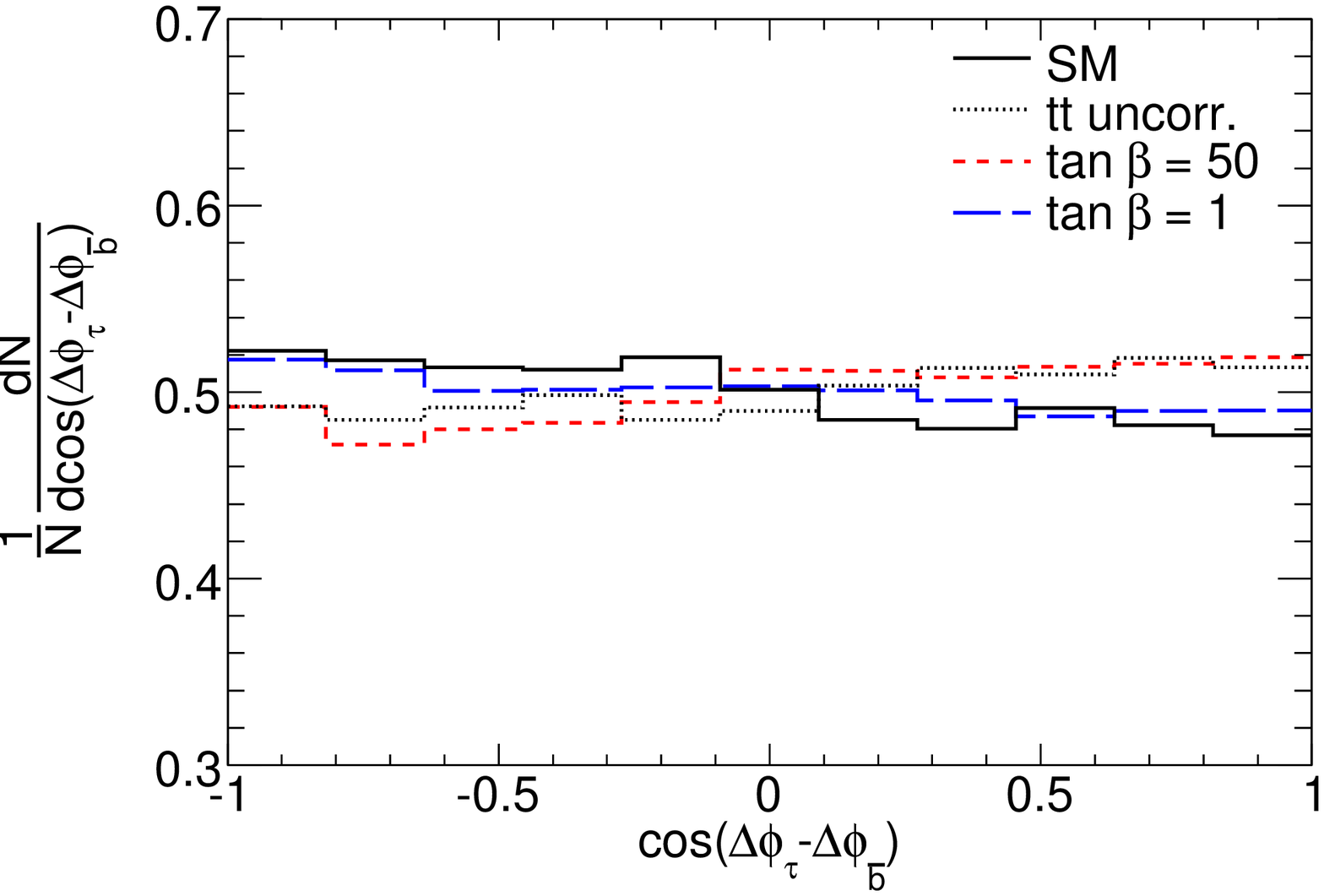}

}
\subfigure{
   \includegraphics[width=0.47\columnwidth,keepaspectratio]{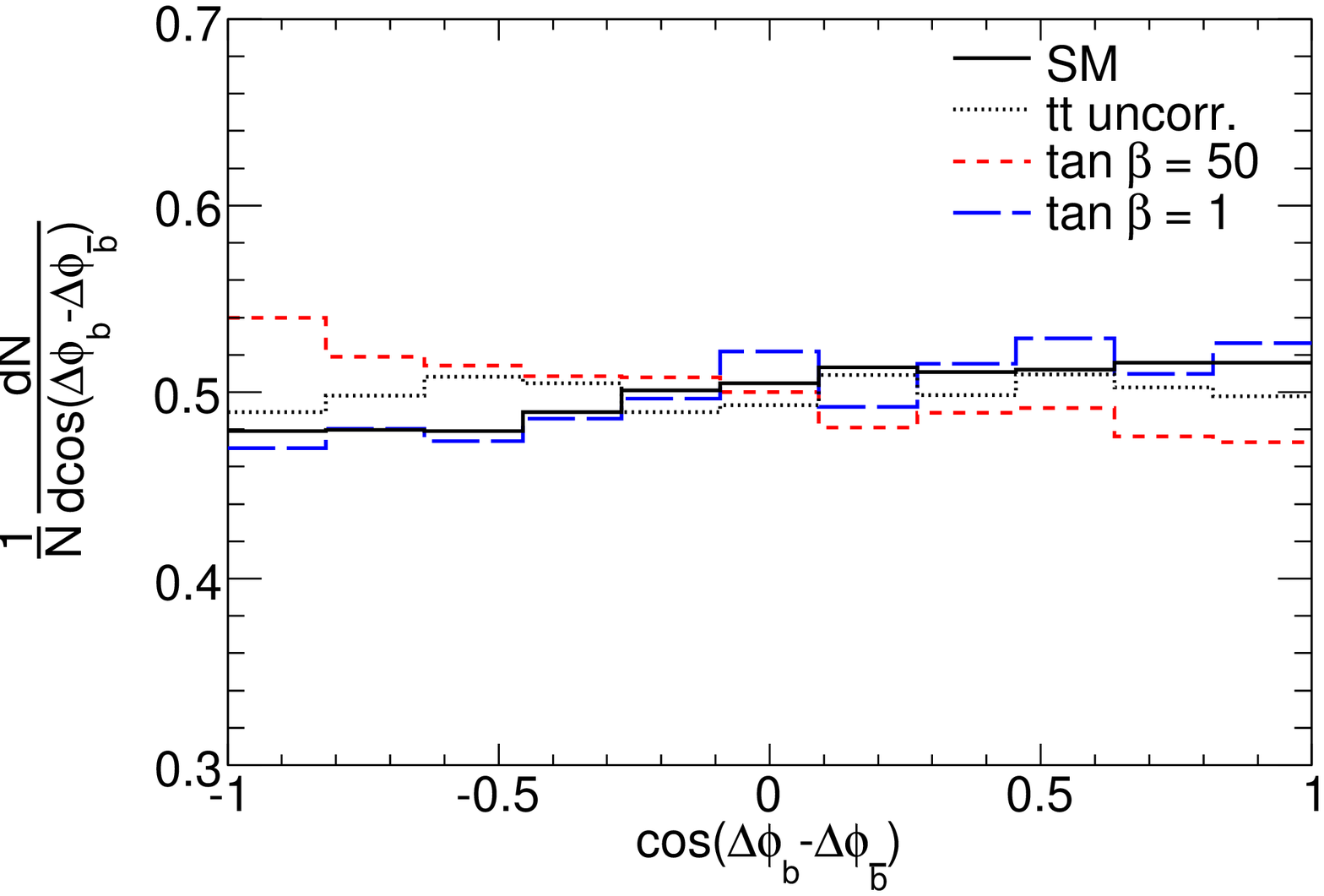}
}
\caption{Jet level distributions for different final state jets $(i,j)$. Colour coding as in fig.~3. Results obtained for $m_{H^+}=80$ GeV. All distributions are normalized individually.}
\label{fig:dphi_jet}
\end{centering}
\end{figure}

Unfortunately, neither of the most sensitive distributions, presented in fig.~\ref{fig:dphill}, are experimentally accessible. This is because there is no lepton originating from the SM side of the event, and the purity in identifying the down-type quark from $W\to q\bar{q}^\prime$ is poor. Hence the less efficient $b$ jet $(|\alpha_b^{\mathrm{SM}}|=0.39$) must be used. In fig.~\ref{fig:dphi_leff} we show the analytic form of the two accessible distributions $(\tau,b)$ and $(b,\bar{b})$. From the MC events we can evaluate the same observables at the hadron level. The result is given in fig.~\ref{fig:dphi_jet} for $m_{H^+}=80$ GeV. The choice of using this very low value for $m_{H^+}$ is simply to isolate the genuine spin effects from the kinematic differences. 

Since we choose to work with normalized distributions, the different histograms in fig.~\ref{fig:dphi_jet} are not normalized to any particular branching ratio $t\to bH^+$. The bin-by-bin fluctuations can be taken as a measure of the statistical variations in the number of events corresponding roughly to $10$ fb$^{-1}$, remembering the crude analysis employed. That the azimuthal distributions have the same advantage as the $\mathcal{D}$-type correlations when it comes to acceptance is clearly illustrated by the uncorrelated sample, which still appears essentially flat also after the cuts.

 We see, as before, that the 2HDM (II) with large $\tan\beta$ differs the most from the V-A structure of the SM. Within the statistical variations, it appears difficult to separate between the 2HDM (II) with low $\tan\beta$ and the SM in any of these channels.

\section{Conclusions}
Top quark spin correlations are predicted by the Standard Model, and this prediction will be tested at the LHC. An angular analysis of $\ttbar$ decay products is most efficiently performed in the dilepton channel which has $\alpha_l=1$. Unfortunately, the same analysis cannot be carried out in the $\tau$ channel due to the additional neutrinos in the final state which prevents reconstruction of the $\ttbar$ CM frame. As we have shown, most of the information about the $t\bar{t}$ spin correlations can however be recovered by looking only in the transverse plane. It would be most interesting to investigate the prospects for top spin physics also in the $\tau$ channel using the tools we suggest.

Such an analysis would make it possible to extract information about the coupling structure of a light charged Higgs boson in the $t\to bH^+\to b\tau^+\nu_\tau$ decay mode. The separation in the angular variables we discuss between the top decay within the SM, and in the charged Higgs decay channel, is hardly of a magnitude interesting for separation of signal and background in a discovery phase. Nevertheless, with the good reach to find a light $H^\pm$ with modest amounts of LHC data \cite{ref:TOP08:Chevallier,ref:TOP08:Yumiceva}, the type of measurements we suggest may still well be considered early in comparison to other physics topics beyond the Standard Model.

\section*{Acknowledgments}
I would like to thank the organizers of TOP2008 for the opportunity to present this talk, and for a very nice and stimulating workshop. The collaboration with David Eriksson, Gunnar Ingelman and Johan Rathsman on this project is acknowledged.

\end{document}